\documentclass[twocolumn,english,superscriptaddress,prb,longbibliography,nobibnotes, groupedaddress]{revtex4-1}
\usepackage{xcolor}
\usepackage{bm}
\usepackage{amstext}
\usepackage{amssymb}
\usepackage{graphicx}
\usepackage{commath}
\usepackage{dsfont}
\usepackage{mathtools}
\usepackage{amsmath,mathrsfs}
\usepackage{physics}
\usepackage{braket}
\usepackage{hyperref}
\usepackage{algpseudocode}
\usepackage{algorithm}
\usepackage{comment}
\usepackage{soul}

\begin{document}
\title{Efficient Paths for Local Counterdiabatic Driving}

\author{Stewart Morawetz}
\email{morawetz@bu.edu}
\affiliation{Department of Physics, Boston University, Boston, Massachusetts 02215, USA}

\author{Anatoli Polkovnikov}
\affiliation{Department of Physics, Boston University, Boston, Massachusetts 02215, USA}

%\date{\today}
%%%%%%%%%%%%%%%%%% ABSTRACT %%%%%%%%%%%%%%%%%%
\begin{abstract}

Local counterdiabatic driving (CD) provides a feasible approach for realizing approximate reversible/adiabatic processes like quantum state preparation using only local controls and without demanding excessively long protocol times. However, in many instances getting high accuracy of such CD protocols requires engineering very complicated new controls or pulse sequences. In this work, we describe a systematic method for altering the adiabatic path by adding extra local controls along which performance of local CD protocols is enhanced, both close to and far away from the adiabatic limit. We also identify an iterative procedure to improve the performance of local counterdiabatic driving further without any knowledge of the quantum wavefunction. We then show that these methods provides dramatic improvement in the preparation of non-trivial GHZ ground states of several different spin systems with both short-range and long-range interactions.

\end{abstract}

\maketitle

%%%%%%%%%%%%%%%%%% INTRODUCTION %%%%%%%%%%%%%%%%%%
\section{Introduction} \label{sec:introduction}

With the rapid development of quantum technologies such as quantum computing, as well as the demands of modern experiments with e.g. cold atoms \cite{vanFrank2016}, trapped ions \cite{Senko2015} and nitrogen-vacancy centers \cite{Rembold2020}, precise control over quantum states is essential. Adiabatic processes present a powerful tool for manipulating these states, where any changes to the system are made sufficiently slowly so that the quantum state remains in an instantaneous eigenstate at all times, allowing for precise control.

However, even in the finite-size systems accessible to near-term quantum computers and modern experimental setups, the timescales required for adiabaticity are often longer than the system remains coherent and thus forbid its application. This has motivated the development of so-called ``Shortcuts to adiabaticity" \cite{Guery-Odelin2019} in which the adiabatic path may be approximately or exactly followed on shorter timescales, at the expense of demanding additional control over the system.

One such technique is known as \textit{counterdiabatic} (CD) \textit{driving} or equivalently transitionless driving~\cite{Demirplak2003, Demirplak2005, Berry2009, Wurtz2022, Chandarana2022, Schindler2023, lawrence2024}, where an additional \textit{counterdiabatic term} is added to the Hamiltonian of the system, which exactly suppresses any transitions between states arising from a fast change of parameters. With this modified Hamiltonian the initial quantum state, which evolves according to the Schr\"odinger equation, will now follow the instantaneous eigenstates by construction.

While this procedure is always exact, the CD term in general requires knowledge of the full spectrum of the system and is thus not accessible for generic many-body systems. This has lead to the development of \textit{local CD driving} \cite{Sels2017, Claeys2019} in which the CD term is restricted to only local operators at the price of failing to completely suppress all transitions. Nonetheless, it yields protocols which can significantly increase the fidelity of the prepared quantum state while remaining feasible to actually implement.

Recently, it has been shown that by combining approaches from quantum optimal control \cite{Stefanatos2020}, whereby additional control terms are added to the system, and then performing local CD driving, quantum states may be prepared with higher fidelity than local CD driving alone \cite{Cepaite2023}. However, it is not immediately clear what these additional control terms should be.

\begin{figure}[h]
    \centering
    \includegraphics[width=.95\linewidth]{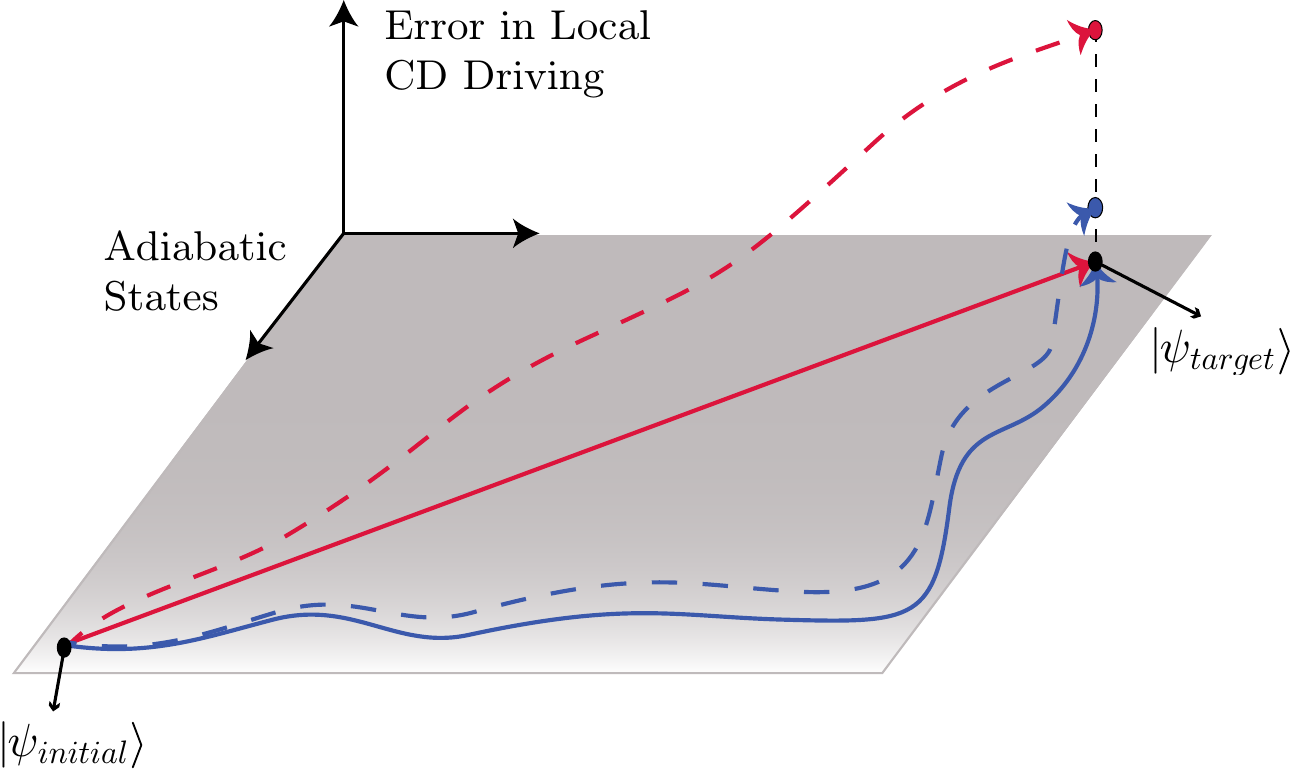}
    \caption{A sketch of the method for finding more efficient paths connecting an initial and the target state. We seek to improve the local CD protocol by adding extra control terms, modifying the adiabatic paths (solid lines) connecting these states  (see also Ref.~\onlinecite{Cepaite2023}). We find that in the situations we analyzed one can drastically improve performance of the local CD protocols by following the optimal path. This is schematically illustrated by a much smaller error for an optimal path (blue) than the ``naive'' path (red) of the original annealing Hamiltonian when performing local CD driving (dashed lines).}
    \label{fig:path-cartoon}
\end{figure}

The main purpose of this work is to propose a systematic method for adding such extra local controls to the Hamiltonian so that local CD protocols are most efficient. Schematically the idea is sketched conceptually in Figure \ref{fig:path-cartoon}. The horizontal plane represents the space of adiabatically connected ground states. By introducing extra controls we can modify the adiabatic path connecting the initial and final states (solid lines). The vertical line schematically represents an error (e.g. deviation of the fidelity from unity) resulting from the local CD driving (dashed lines). The optimal (blue) path results in a lower error. Note that while we focus on quantum state preparation this formalism can be applied to realize fast and reversible energy transfer to facilitate various thermodynamic processes~\cite{Villazon_2019,Gjonbalaj2022}.

This paper is organized as follows: in Section \ref{sec:adiabatic-state-prep} we review how quantum states may be prepared adiabatically, when this fails, and how counterdiabatic driving may be employed to alleviate this. In Section \ref{sec:extra-controls} we discuss how we may employ local (approximate) counterdiabatic driving more efficiently by augmenting the underlying Hamiltonian with extra control terms. In Section \ref{sec:results} we show how this may be applied beyond the standard short-range models, especially those connected to relevant experiments. We also identify a novel iterative procedure for further improving the local CD driving protocol in the absence of extra controls. Finally, we summarize in Section \ref{sec:conclusions} and discuss potentially fruitful applications of these techniques beyond the contents of this paper.

%%%%%%%%%%%%%%%%%% ADIABATIC STATE PREPARATION %%%%%%%%%%%%%%%%%%
\section{Counterdiabatic driving \label{sec:adiabatic-state-prep}}

One straightforward approach to preparing quantum states is to employ the adiabatic theorem. For a system whose energy levels have a finite gap, this guarantees \cite{Sakurai2017} that if the control parameter of the Hamiltonian is changed sufficiently slowly, the system will always remain in an instantaneous eigenstate.

Let us consider a time dependent Hamiltonian $H(\lambda)$, where the time dependence is encoded in a control parameter $\lambda = \lambda(t)$ which varies from 0 to 1. While this may in general represent a vector of several control parameters, we will restrict ourselves to just one for simplicity. The adiabatic theorem guarantees that if our system begins in the ground state of $H(\lambda = 0)$, and the parameter $\lambda$ is changed sufficiently slowly, the system will conclude the protocol in the ground state of $H(\lambda = 1)$.

One straightforward application of this is so-called quantum annealing, where the Hamiltonian is designed such that the final ground state at $\lambda = 1$ is the target state, which is usually difficult to prepare, but the initial ground state at $\lambda = 0$ is easy to prepare. By preparing the system in the initial state and then changing $\lambda$ slowly, the target state may be obtained with arbitrarily high fidelity. However, the condition that $|\dot{\lambda}|$ be small can become quite restrictive as we move beyond simple systems and the gaps between energy levels become smaller.

Formally, the adiabatic approximation is made by writing the Schr\"odinger equation in the basis of instantaneous eigenstates, and then neglecting emergent rate dependent off-diagonal terms coupling different eigenstates~\cite{Berry2009,Kolodrubetz2017}. This suggests that perfect adiabaticity may be obtained by inserting into the Hamiltonian some compensating terms which will exactly cancel those ignored in the adiabatic approximation. If this is done, the adiabatic approximation becomes exact irrespective of $|\dot\lambda|$.

Counterdiabatic driving was first discovered by Demirplak \& Rice \cite{Demirplak2003, Demirplak2005} in the context of population transfer between molecular states, and independently formulated by Berry \cite{Berry2009} and termed ``transitionless driving,'' although the two are entirely equivalent. This involves defining the \textit{counterdiabatic Hamiltonian} $H_{\rm CD}$ such that

\begin{equation} \label{eqn:H_CD}
    H_{\rm CD} = H + \dot{\lambda} A_\lambda
\end{equation}

\noindent where $A_\lambda$ is known as the \textit{adiabatic gauge potential} (AGP). This operator is responsible for transforming the instantaneous eigenstates under a change of the control parameter $\lambda$, and it is the term which makes the adiabatic approximation exact. There are many possible ways to encode $\lambda$ as a function of time. Following earlier work, we encode it in the following smooth way
\begin{equation} \label{eqn:lambda_encoding}
    \lambda(t) = \sin^2 \left( \frac{\pi}{2} \sin^2\left( \frac{\pi t}{2 \tau} \right) \right)
\end{equation}
\noindent so that $\lambda(0) = 0$ and $\lambda(\tau) = 1$ . The ``speed'' of the process is encoded in $\tau$. Since $\dot{\lambda} \propto 1/\tau$, when $\tau$ is small, the counterdiabatic term $\dot{\lambda} A_\lambda$ dominates the dynamics defined by $H_{\rm CD}$ in Eq. \eqref{eqn:H_CD}. As $\tau$ is increased, this term vanishes and the process becomes adiabatic.

Formally, the AGP satisfies the following equation (in units where $\hbar = 1$)

\begin{equation} \label{eqn:AGP_commutator_eqn}
    [\partial_\lambda H + i [A_\lambda, H], H] = 0
\end{equation}

\noindent Despite its apparent simplicity, this equation is in general very difficult to solve. It can be shown \cite{Sels2017} however that solving Eq. \eqref{eqn:AGP_commutator_eqn} is equivalent to minimizing the following action:

\begin{equation} \label{eqn:variational_action}
    S_\lambda(A_\lambda) = \operatorname{Tr} \left[ G_\lambda^2 \right];
    \ G_\lambda = \partial_\lambda H + i [A_\lambda, H].
\end{equation}

\noindent We note in passing that one can use replace $\operatorname{Tr} \left[ G_\lambda^2 \right]\to \operatorname{Tr} \left[\rho\, G_\lambda^2 \right]$, where $\rho$ is an arbitrary stationary density matrix with respect to the Hamiltonian $H(\lambda)$~\cite{Sels2017}. By using $\rho = {1\over Z}\exp{-\beta H}$, we can perform a finite-temperature variational optimization, which preferentially targets lower-energy eigenstates and can give better performance in some instances. In the limit $\beta\to \infty$ the average with respect to $\rho$ reduces to the ground state expectation value. In Sec.~\ref{subsec:finite-temp} we discuss in detail how one can further improve local CD protocols using the ground state optimization of the AGP without requiring any prior knowledge of the ground state. We may interpret Eq.~\eqref{eqn:AGP_commutator_eqn}, which in this language reads $[G_\lambda, H] = 0$, as a statement  that a well-defined AGP $A_\lambda$  admits the existence of a conserved operator $G_\lambda$ commuting with $H$.

Instead of dealing with the exact AGP, which is very nonlocal and even ill-defined in chaotic systems~\cite{Pandey2020}, it is convenient to find an approximate local AGP, which minimizes the action in Eq.~\eqref{eqn:variational_action} within a restricted subset of operators $A_\lambda$. A very convenient option is to choose this subset from the so-called Krylov space, which is obtained by a repeated action of Liouvillian $\mathcal{L} = [H, \cdot]$ on $\partial_\lambda H$~\cite{Claeys2019}:

\begin{equation} \label{eqn:commutator_ansatz}
    A_\lambda^{(\ell)} = i \sum_{k = 1}^\ell \alpha_k \underbrace{[H,[H, ..., [H}_{2k-1}, \partial_\lambda H]]].
\end{equation}

Notably, only odd orders of nested commutators enter into this ansatz. Here $\ell$ controls the order (and therefore the locality) of the expansion, and $\alpha_k$ are the variational parameters found by minimizing the action~\eqref{eqn:variational_action}. Formally this ansatz is exact in the limit $\ell\to \infty$ but in practice we want to restrict it to small values of $\ell$. This process can by made more numerically stable by using the Lanczos algorithm in the operator basis leading to an AGP expansion in an orthonormal set of Krylov operators $O_k$ (see Refs.~\onlinecite{Takahashi2023, Bhattacharjee2023,Sels_2023} for details). We give a detailed description of our Krylov space construction of the AGP in Appendix \ref{appendix:krylov_construction}.
 %%%%%%%%%%%%%%%%%% FINDING EXTRA CONTROLS %%%%%%%%%%%%%%%%%%
\section{Improving CD driving with extra controls} \label{sec:extra-controls}

The main goal of this section and of the whole paper is to find a systematic approach to improving performance of approximate CD driving via adding extra controls to the Hamiltonian. We test these by applying them to prepare nontrivial GHZ entangled states. We will begin by focusing on the infinite speed limit $|\dot\lambda|\to\infty$, where although the $H$ term in \eqref{eqn:H_CD} is included in the simulations, the evolution according to $H_{\rm CD}$ is dominated by the counterdiabatic term $\dot{\lambda} A_\lambda$. In \ref{subsec:finite_time}, we discuss the improvement from the extra controls in the regime where $\dot{\lambda}$ is finite, when both $H$ and $A_\lambda$ play a meaningful role in the dynamics.

Before proceeding with nontrivial systems, we illustrate the idea of extra controls using an intuitive example of ground state preparation in a one-dimensional Ising model with transverse and longitudinal fields described by a standard quantum annealing scheme~\cite{Hauke_2020}:

\begin{multline}
\label{eq:annealing}
    H(\lambda) = \lambda H_0+(1-\lambda) H_1,\\
    H_0=-\sum_i \sigma_i^z \sigma_{i+1}^z + h_z \sigma_i^z+h_x \sigma_i^x,\quad H_1= - \sum_i \sigma_i^x
\end{multline}

\noindent with periodic boundary conditions. We will start from the situation where there is a small but finite longitudinal field $0<h_z\ll 1$ and finite $h_x<1$. The first condition breaks the $Z_2$ symmetry of the model such that the ferromagnetic ground state for $|h_x|<1$ is polarized in the positive $z$-direction and the second condition implies that during the annealing protocol the system crosses a quantum phase transition at $h_x=1$~\cite{sachdev2011}. At this point the gap closes and the AGP becomes an operator with an infinite range~\cite{del_Campo_2012, Damski_2014,Kolodrubetz2017} such that local CD protocols become very inefficient.

\begin{figure}[h]
    \centering
    \includegraphics[width=\linewidth]{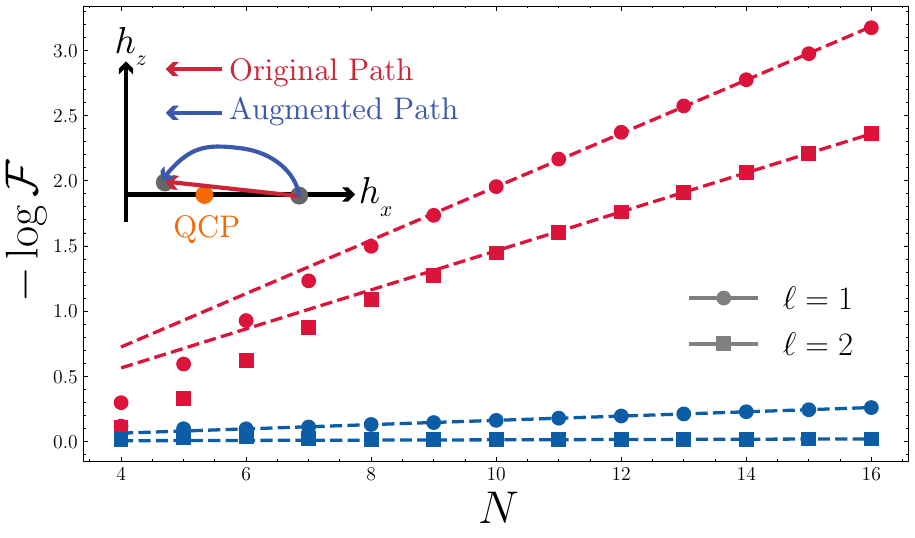}
    \caption{A comparison between the results when performing local CD driving on the ``naive'' original path in red, which passes close by the critical point at $(h_x, h_z) = (1, 0)$, and the augmented path in blue, where the critical point is avoided. The final Hamiltonian $H_0$ is given by Eq.~\eqref{eq:annealing} with $h_x=0.7,\; h_z=0.01$. The circles and squares show the results using $\ell = 1$ and $\ell = 2$ in Eq. \eqref{eqn:commutator_ansatz} respectively. We restrict to $\beta \leq 3$ as defined in Eq. \eqref{eqn:polarized_extra_terms} to ensure that we are not finding optimal paths by taking $\beta$ large and effectively rescaling time.}
    \label{fig:LTFIM_sketch}
\end{figure}

In this setup it should be clearly more advantageous to choose an alternate path, which while retaining the same start and end points, stays far away from the critical point. This is shown in the inset of Figure \ref{fig:LTFIM_sketch}. Following the general idea of Ref.~\onlinecite{Cepaite2023} the Hamiltonian corresponding to the modified path is given by
\begin{equation} \label{eqn:polarized_extra_terms}
    \tilde{H}(\lambda) = H(\lambda) + \beta \sin(\pi \lambda) \sum_i \sigma_i^z,
\end{equation}

\noindent where $\beta$ is some parameter which we can optimize numerically. We then apply local CD driving to $\tilde{H}(\lambda)$. With an appropriate choice of $\beta$, this gives a very strong improvement in state preparation, as shown in Figure \ref{fig:LTFIM_sketch}. In particular, we can see that even just the first two terms in the expansion in Equation \eqref{eqn:commutator_ansatz} are sufficient to prepare the state with high fidelity nearly independent of system size despite the fact that $h_x=0.7$ is not very small and thus the ferromagnetic state is far from fully polarized.

\subsection{Formulation of General Problem} \label{subsec:extra-controls-general-problem}

The example above shows the utility of adding extra controls, which in that case can be intuitively found. Similar ideas without CD driving were experimentally implemented to prepare topological Hofstadter bands in ultracold atoms~\cite{Aidelsburger_2014}. The question we aim to address is how can we find efficient extra controls for arbitrary Hamiltonians $H_0$ and $H_1$. Formally we define an alternate path by evolving the system according to an \textit{augmented Hamiltonian} 
\begin{equation} \label{eqn:aug_Hamiltonian}
    \tilde{H}(\lambda) = H(\lambda) + H_c(\lambda),\quad H_c(\lambda)=\sum_n \beta^{(n)}(\lambda) H_c^{(n)}
\end{equation}

\noindent where $H(\lambda$) is given by Eq.~\eqref{eq:annealing} and the $\beta^{(n)}(\lambda)$ are some smooth functions satisfying boundary conditions $\beta^{(n)}(\lambda = 0) = \beta^{(n)}(\lambda = 1) = 0$ so that the initial and final eigenstates of the annealing Hamiltonian are unchanged. We refer to $H_c$ as an \textit{extra control Hamiltonian}. In the previous example, it was just an additional global $z$ field, i.e. $H_c = \sum_i \sigma_i^z$. We stress that in the infinite speed limit the dynamics is determined exclusively by the AGP, so the annealing Hamiltonian plays a purely auxiliary role like e.g. in Ref.~\onlinecite{Ljubotina_2022}.

We then perform local CD driving for the augmented Hamiltonian $\tilde{H}(\lambda)$, hence evolving according to $\tilde{H}_{CD}(\lambda)$.The goal is to choose both $\beta^{(n)}(\lambda)$ and $H_c^{(n)}$ to maximize the fidelity of the final state. The $\beta^{(n)}(\lambda)$ are fixed for a given evolution, and the variational parameters $\alpha_k$ of the AGP are determined locally at each time step. Finding the optimal control functions $\beta^{(n)}(\lambda)$ is the subject of quantum optimal control \cite{Glaser2015} or similar methods \cite{Bukov2018}, which is not the main focus of this work. We pick the $\beta^{(n)}(\lambda)$ by a very simple optimization, following earlier work \cite{Cepaite2023}, writing it as a single harmonic term

\begin{equation} \label{eqn:beta}
    \beta^{(n)}(\lambda) = \beta^{(n)} \sin(\pi \lambda)
\end{equation}

\noindent and choosing  $\beta^{(n)}$ to maximize the fidelity of the final state. This is done by maximizing the fidelity

\begin{equation} \label{eqn:fidelity}
    \mathcal{F}(\beta) = \vert \braket{\psi_{target} | \psi_{evolved}(\beta)} \vert^2
\end{equation}

\noindent where $\ket{\psi_{evolved}(\beta)}$ is obtained by evolving the initial state by the augmented CD Hamiltonian $\tilde{H}_{CD}$. We note that the fidelities can be even further improved by adding additional couplings $\beta_k^{(n)}$ corresponding to $k$-th harmonic of $\pi\lambda$~\cite{Cepaite2023}. We limit ourselves only to the first one as we want to focus on the question of finding optimal $H_c^{(n)}$.

\subsection{Ansatz for Extra Controls} \label{subsec:extra-controls-ansatz}

To proceed, let us consider a state preparation problem similar to the previous one, but with $h_x = h_z = 0$, i.e. without breaking the $Z_2$-symmetry. The ferromagnetic ground state of this model is now a GHZ state \cite{Greenberger1989}. It is defined as an equal superposition of all spins up and all spins down. This adiabatic state preparation is encoded in the Hamiltonian 
\begin{equation} \label{eqn:short-range}
    H_0=-\sum_i \sigma_i^z \sigma_{i+1}^z,\quad H_1= - \sum_i \sigma_i^x
\end{equation}
The previous choice of $H_c = \sum_i \sigma_i^z$ will not work, since it breaks the $Z_2$ symmetry of the ground state of $H_0$ and allows one to prepare either the up or down polarized state, but not a superposition of both.

In order to identify the form of $H_c$, we note that for any value of $\lambda$ the AGP ansatz in Equation \eqref{eqn:commutator_ansatz} is composed of odd commutators of the operators $H_0$ and $H_1$ like
\[
i[H_1,H_0],\; i [H_0,[H_0,[H_1,H_0]]],\;
i [H_0,[H_1,[H_1,H_0]]],\dots
\]
In the AGP these commutators appear with different coefficients which depend on $\lambda$. Let us observe that the general composition of the operators entering the exact AGP (i.e. at all orders of the expansion in Eq. \ref{eqn:commutator_ansatz}) will not change if we modify the Hamiltonian $H(\lambda)$ by adding arbitrary \textit{even} order commutators to it, i.e. adding terms like
\begin{equation}
\label{eq:extra_even}
H_c^{(1)}=[H_0,[H_1,H_0]],\quad H_c^{(2)}=[H_1,[H_1,H_0]],\dots 
\end{equation}
with the corresponding weights $\beta^{(n)}(\lambda)$ as in Eq. \eqref{eqn:aug_Hamiltonian}. By adding such terms we clearly expand the variational manifold allowing one to gradually increase the locality of both $\tilde H(\lambda)$ and $A_\lambda$. Moreover such terms can be generated using Floquet pulse sequences containing only $H_0$ and $H_1$ and is routinely done in the NMR literature~\cite{farrar1971pulse}. A detailed description of how to implement a version of our protocol with Floquet pulses is given in Appendix \ref{appendix:floquet}. A very similar approach is described in \onlinecite{Claeys2019}. As we show below, for the examples we analyze here this idea leads to the dramatic improvement of local CD protocols.

It is easy to check that for the Ising model the new terms (i.e. terms which are not originally present in $H_0$) which appear by computing the commutators in Eq.~\eqref{eq:extra_even} are $YY$ and $ZXZ$, where we use a common short-hand notation: $YY=\sum_j \sigma_j^y \sigma_{j+1}^y$, $ZXZ=\sum_j \sigma_j^z \sigma_{j+1}^x \sigma_{j+2}^z$. We can thus select $H_c^{(1)} = YY$ and $H_c^{(2)} = ZXZ$. In this way the augmented Hamiltonian~\eqref{eqn:aug_Hamiltonian} will cause the state to follow a genuinely new path. The resulting improvement of the final state fidelity is plotted in Figure \ref{fig:local_model_yy_zxz_results}. For a more detailed discussion of the optimization procedure see Appendix \ref{appendix:YY_vs_ZXZ}.

\begin{figure}
    \centering
    \includegraphics[width=\linewidth]{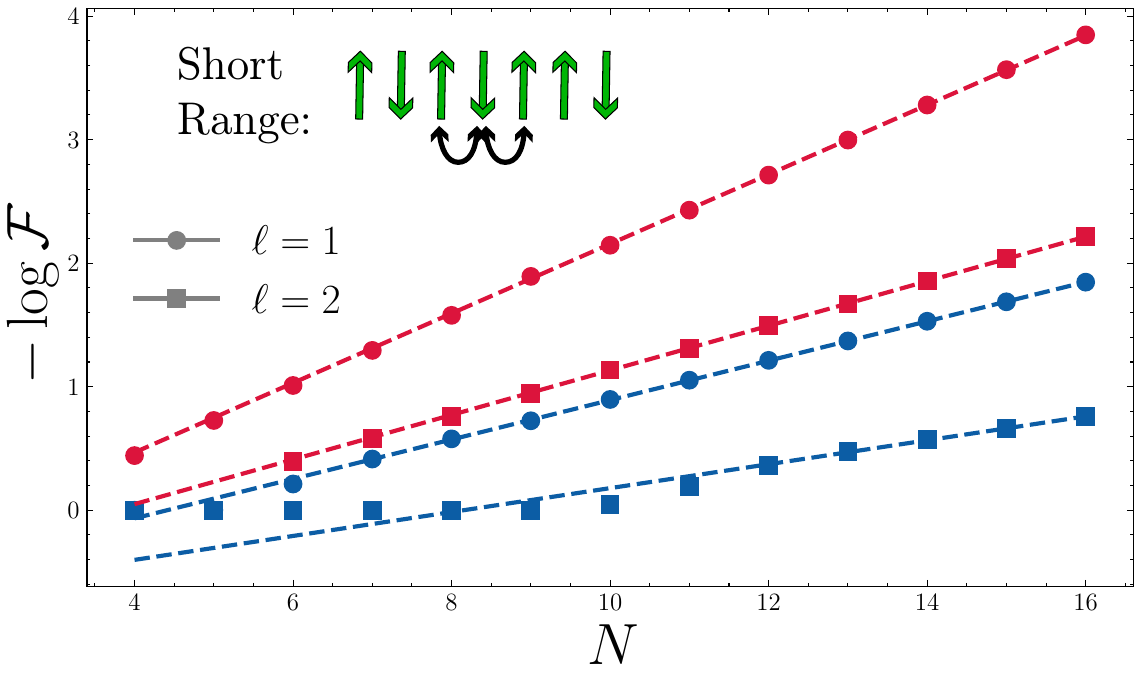}
    \caption{Improvement in the final state fidelity when prepared with local counterdiabatic driving. In red, we perform local CD driving along the original path given by Eq. \eqref{eq:annealing}, whereas in blue it is performed along a path where we augment Eq. \eqref{eq:annealing} by $H_c^{(1)} = YY$ and $H_c^{(2)} = ZXZ$ as in \eqref{eq:extra_even}. The circles and squares of red and blue lines correspond to $\ell=1$ and $\ell=2$ in Eq.~\eqref{eqn:commutator_ansatz} respectively.}
    \label{fig:local_model_yy_zxz_results}
\end{figure}

We can see that the extra controls lead to a dramatic improvement of the protocol. In particular, for system sizes up to $N=9$ the second order AGP ansatz corresponding to $\ell=2$ in Eq.~\eqref{eqn:commutator_ansatz} leads to nearly unit fidelity. For larger system sizes the fidelity decays exponentially with $N$ with a reduced slope, and thus still offering exponential enhancement over the original CD protocol. We note that as we continue to increase the order of the AGP expansion the size of the system we can prepare with unit fidelity increases, as does the exponential enhancement in fidelity over the corresponding original CD protocol. 

\subsection{Finite-Time Protocol Performance} \label{subsec:finite_time}

Up to this point, we have only considered the case where $|\dot\lambda|\to\infty$ and the dynamics are controlled entirely by the AGP $A_\lambda$ in Eq. \eqref{eqn:H_CD}. There is also the trivial adiabatic limit, where $|\dot{\lambda}|\to0$ and the eigenstates are transported perfectly with the AGP playing no role. But in between these, there is a regime where both the Hamiltonian and the AGP play an important role in the counterdiabatic dynamics.

\begin{figure}
    \centering
    \includegraphics[width=\linewidth]{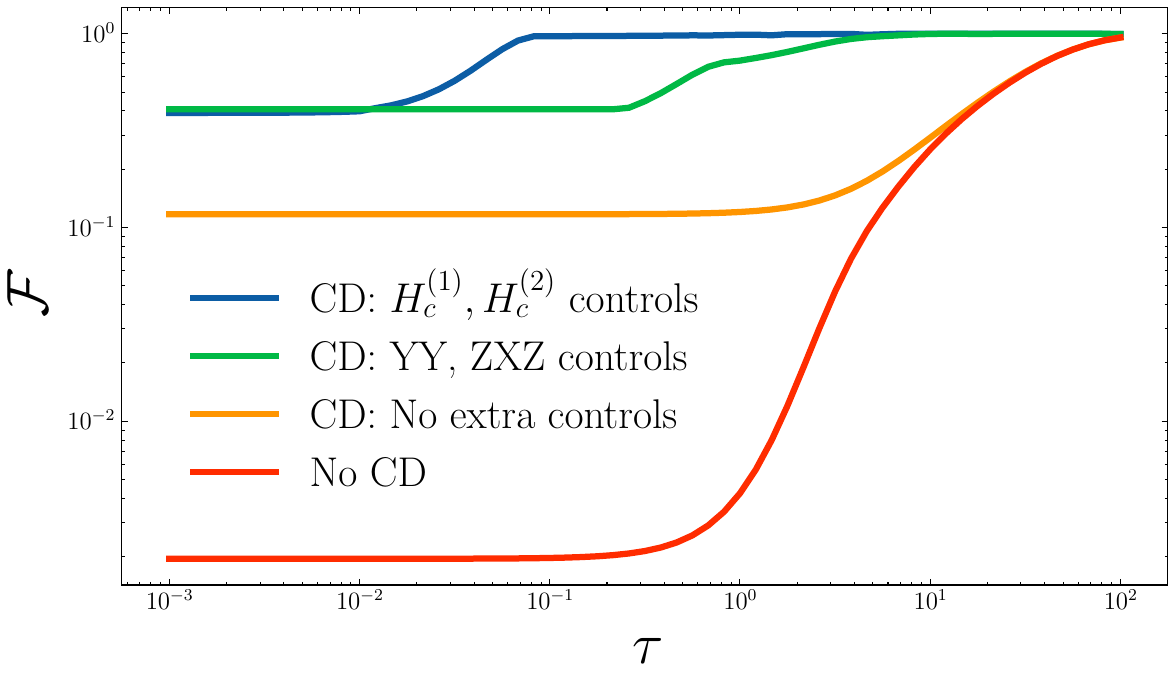}
    \caption{The final state fidelity for the annealing problem of Eq. \eqref{eqn:short-range} with $N = 8$, for different total protocol times. The final state fidelity is improved not only in the ``fast'' limit with $\tau\to0$, but for any intermediate protocol duration up to the adiabatic limit. Particularly noteworthy is that even local controls are sufficient to reduce the threshold for adiabatic evolution by several orders of magnitude.}
    \label{fig:finite-time}
\end{figure}

For the same annealing problem defined in Eq. \eqref{eqn:short-range}, we plot the final state fidelity when we perform CD driving with different control schemes in Figure \ref{fig:finite-time}. From this, one can see that different extra control schemes offer improvement in the finite time regime, in addition to the $|\dot{\lambda}|\to\infty$ ($\tau\to0$) limit discussed earlier. In particular, using the recipe for extra controls given in Eq. \eqref{eq:extra_even}, one can prepare states with unit fidelity in a time reduced by nearly three orders of magnitude.

%%%%%%%%%%%%%%%%%% RESULTS %%%%%%%%%%%%%%%%%%
\section{Preparing GHZ state in a Long range model} \label{sec:results}

\subsection{Fidelity Gain from Augmented Hamiltonian} \label{subsec:long-range-recipe}

We now move to preparation of the GHZ state in longer range Ising Hamiltonians, which are relevant to cold atom/trapped ion systems \cite{Koffel2012, Jaschke2017, Adelhardt2020}. We define the annealing Hamiltonian for a long-range Ising model:
\begin{equation} \label{eqn:LR_ising}
    H_0 = -\sum_{i,j} \frac{1}{\vert i - j\vert^\alpha} \sigma_i^z \sigma_j^z,\quad H_1= - \sum_i \sigma_i^x,
\end{equation}
where $\alpha>0$. Note that we can interpolate between the short-range model studied earlier and a fully connected model by varying the exponent $\alpha$.  

Let us first consider a long-range case with $\alpha=2$ and then move to the fully connected model with $\alpha=0$. In both cases the naive path crosses a critical (gapless) point at some value of $\lambda$ which depends on the parameter $\alpha$ and additionally scales with $N$ for $\alpha\leq 1$.

We then augment the original Hamiltonian by introducing extra controls
\[
\beta^{(1)}(\lambda)=\beta^{(1)} \sin(\pi\lambda),\quad \beta^{(2)}(\lambda)=\beta^{(2)} \sin(\pi\lambda)
\]
corresponding to the terms $H_c^{(1)}$ and $H_c^{(2)}$ in Eq.~\eqref{eq:extra_even}. Like in the previous example we then use a standard CD protocol and find the parameters $\beta^{(1)}$ and $\beta^{(2)}$ by numerically optimizing the fidelity of the final state.

\begin{figure}
    \centering
    \includegraphics[width=\linewidth]{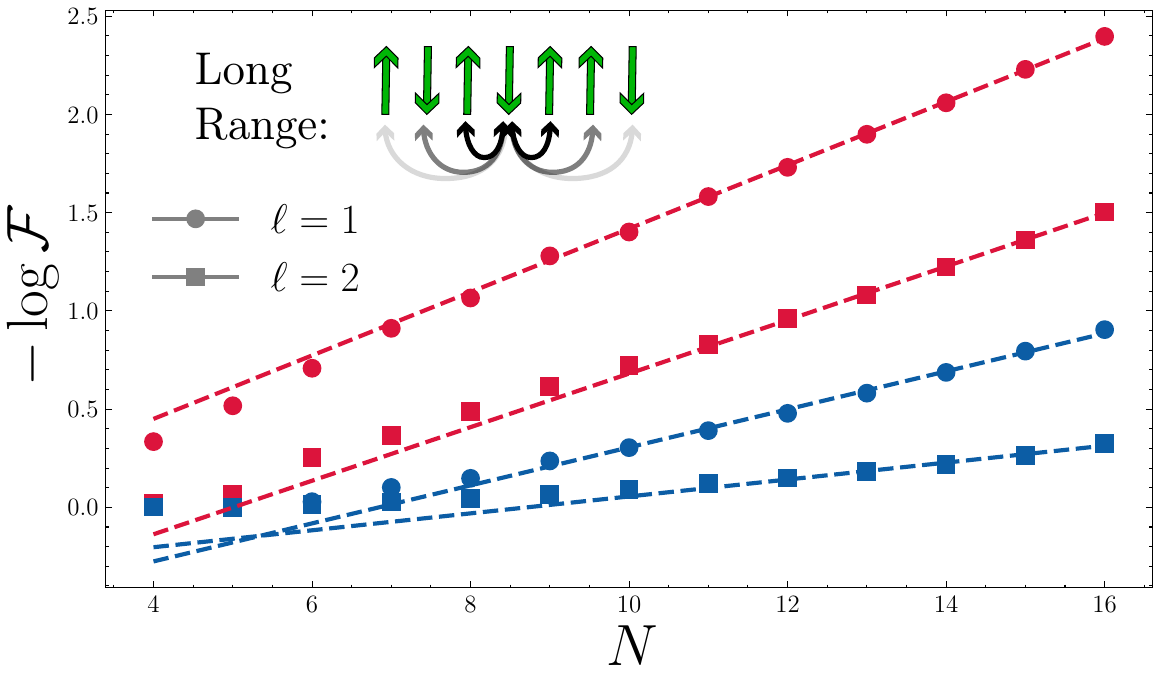}
    \caption{Improvement in fidelity obtained by a more efficient path for local CD driving with the intermediate-range Ising model of Eq. \eqref{eqn:LR_ising}. As before, in red we show results for local CD driving on the original Hamiltonian, and in blue for the Hamiltonian augmented by extra terms of the form \eqref{eq:extra_even}. The two sets of lines correspond to $\ell = 1$ and $\ell = 2$ in Eq. \eqref{eqn:commutator_ansatz}.}
    \label{fig:LR_results}
\end{figure}

In Figure ~\ref{fig:LR_results}, for the long-range Ising model, we find a dramatic improvement in the GHZ state preparation along the optimally augmented path. We notice that this improvement is not tied in any way to the integrability (short range)/nonintegrability (long range) of the Ising model. This is not that surprising, as the short range AGP (small $\ell$) cannot distinguish integrable and nonintegrable systems~\cite{Pandey2020} and extra controls generally break integrability anyway.

\begin{figure}
    \centering
    \includegraphics[width=\linewidth]{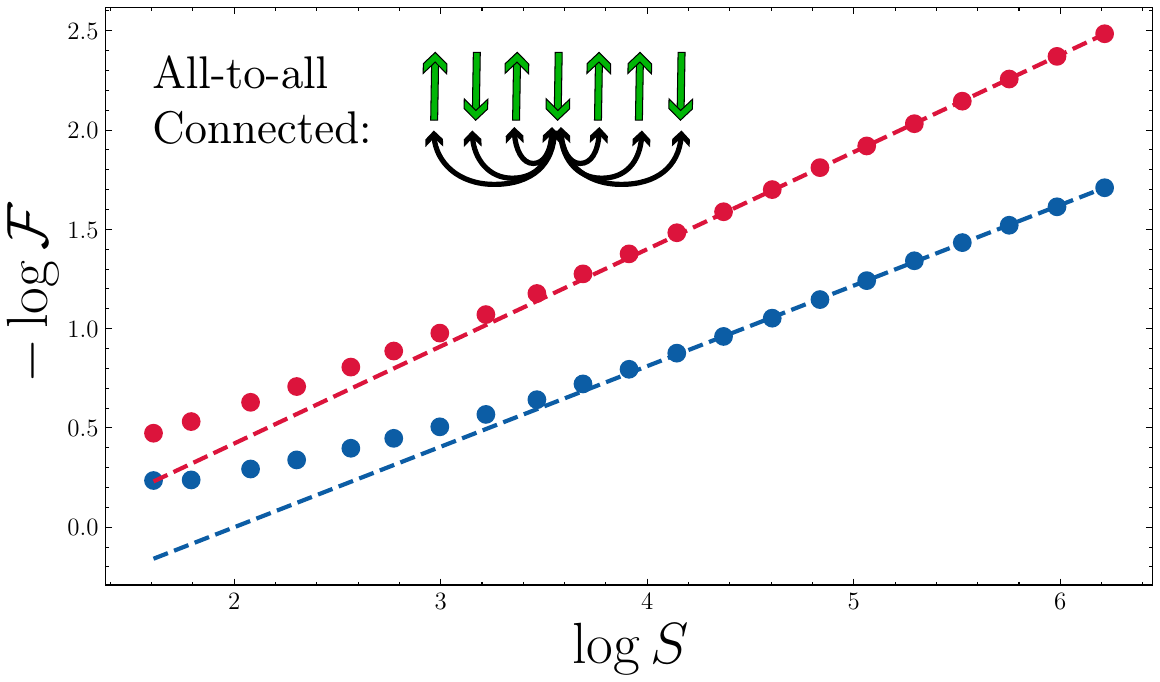}
    \caption{Restricting to only the first order term in the expansion of Eq. \eqref{eqn:commutator_ansatz}, we perform local CD driving on both the original (red) spin-squeezing Hamiltonian of Eq. \eqref{eqn:spin-squeezing}, and where it is augmented by the extra controls ansatz (blue). We find that improvement in the final state fidelity using the extra controls Hamiltonian Eq. \eqref{eq:extra_even}.}
    \label{fig:spin_squeezing}
\end{figure}

Finally, let us discuss the fully connected model corresponding to $\alpha=0$. With an appropriate rescaling of couplings, this model is equivalent to a single large spin $\vec S=\frac{1}{2}\sum_j \vec \sigma_j$ with a nonlinear interaction. The corresponding annealing Hamiltonian is still given by Eq.~\eqref{eq:annealing} with
\begin{equation} \label{eqn:spin-squeezing}
    H_0 = \frac{1}{\sqrt{S(S+1)}} S_z^2,\quad  H_1= S_x.
\end{equation}
Here we rescale the first term so that the model has a well defined classical/thermodynamic limit as $S=N/2\to\infty$. We note in passing that this Hamiltonian is extensively employed for experimental preparation of spin squeezed states~\cite{Kitagawa1993,law2001,rojo2003,Sorensen2001}. Such spin-squeezed states allow for better scaling of measurement precision in Ramsey interferometry, surpassing the Standard Quantum Limit \cite{Wineland1992, Bollinger1996}. The results for GHZ state preparation with this Hamiltonian are shown in Figure \ref{fig:spin_squeezing}. Interestingly one of the emergent extra control Hamiltonians in Eq. \ref{eq:extra_even} is $S_y^2$, which is used in the two-axis twisting Hamiltonian for the preparation of even better spin-squeezed states.

As in both the short- and long-range spin models, we observe that we can prepare states with much better fidelity using these more efficient paths, though the improvement is less dramatic. The performance could be enhanced further by considering finite energy norms for the variational optimization of the action, as we briefly discuss next and as was done in a different classical model~\cite{Gjonbalaj2022}.

\subsection{Ground State Optimization} \label{subsec:finite-temp}

As previously mentioned, we can replace $\operatorname{Tr}\left[G_\lambda^2\right]$ in Eq. \eqref{eqn:variational_action} by $\operatorname{Tr}\left[ G_\lambda^2 \exp\left(-\beta H\right)\right]$, where $\beta$ is the inverse temperature. The original action is thus equivalent to an infinite temperature ($\beta = 0$) optimization. By changing the temperature, higher weight will be assigned to lower-energy eigenstates in the optimization. As we are concerned in this paper with quantum annealing problems involving the ground state, the natural question to ask is what happens when we use a zero temperature ($\beta = \infty$) optimization, where we \textit{only} optimize over transitions into and out of the ground state. Because the effect of ground state vs. infinite temperature optimization is separate from finding extra controls, in this section we only focus on the local CD driving along the original adiabatic path. One can of course combine the two approaches together, but such full optimization is outside the scope of this work. We note on passing that in some situations like in classical systems one has to deal with finite temperature actions to get meaningful results for the AGP~\cite{Gjonbalaj2022}.

\begin{figure}
    \centering
    \includegraphics[width=\linewidth]{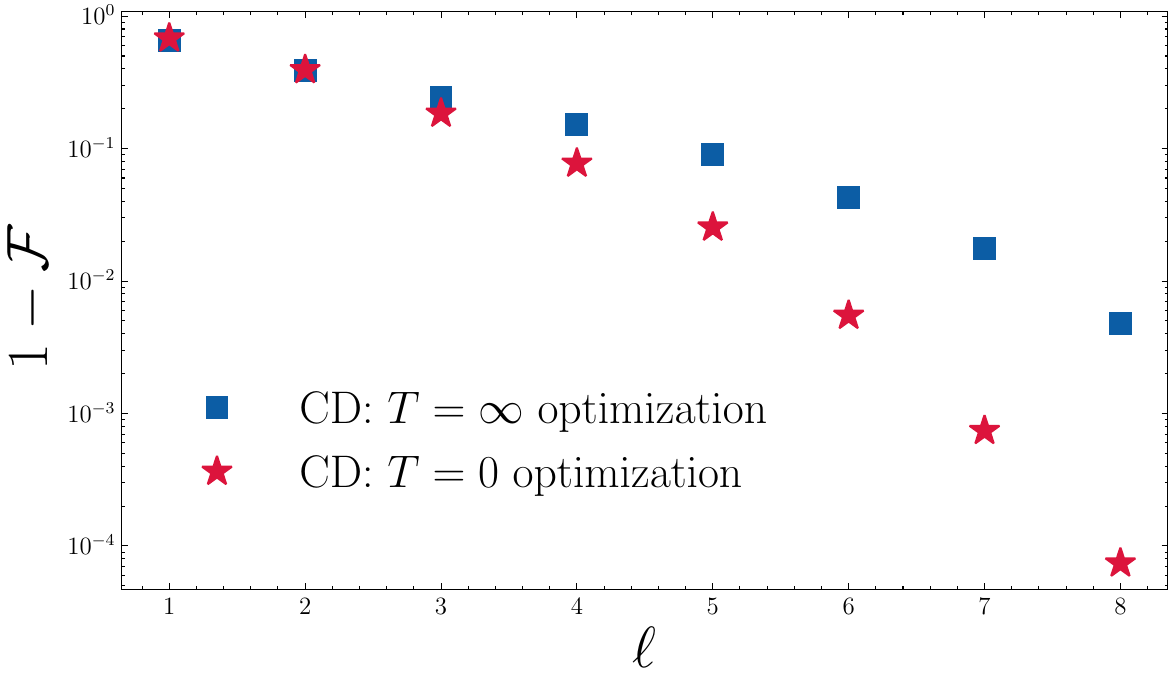}
    \caption{For the annealing problem of Eq. \eqref{eqn:LR_ising} with $N = 8$, we show the difference in final state fidelity between the infinite temperature (all eigenstates) and zero temperature (ground state only) optimization of the AGP. Here, $\ell$ is as defined in Eq. \eqref{eqn:commutator_ansatz}. As the order increases and the evolved state is increasingly constrained to the ground state manifold, the zero temperature optimization becomes superior.}
    \label{fig:optim_temp_compare_order}
\end{figure}

To study the effect of the ground state action, we consider the long-range model with $\alpha = 2$ defined in Eq. \eqref{eqn:LR_ising}. We note that if we use the short range model corresponding to $\alpha=\infty$ instead then there is no effect of temperature on the action. This fact follows from the equivalence of the short-range model to a set of independent  two-level systems, where all operator norms are identical for the excited and ground states~\cite{Kolodrubetz2017}. In Figure \ref{fig:optim_temp_compare_order} we plot the fidelities for the local CD protocols obtained using infinite and zero temperature actions. We see that at sufficiently high orders of the variational ansatz where the infinite temperature protocol already performs very well, the ground state optimization provides even further improvement. However, at small $\ell<3$ the ground state optimization tends to do slightly worse. A possible physical explanation is that the ground state protocol does poorly at suppressing transitions from the excited states, so if the system gets excited the infinite temperature protocol does a better job of suppressing further excitations.

In principle, unlike the $T = \infty$ optimization, the $T = 0$ optimization requires a priori knowledge of the ground state, which would generally require a separate quantum simulator. However, we can use instead the approximate ground state prepared using the infinite-temperature CD protocol to further optimize the local AGP, and then use  that new local AGP to prepare a better ground state. This procedure can be iterated, returning the approximate ground state AGP. We note that this iterative scheme does not require any extra computational resources and can be done using the same simulator (experimental or computational) used to simulate time evolution of the state.

\begin{figure}
    \centering
    \includegraphics[width=\linewidth]{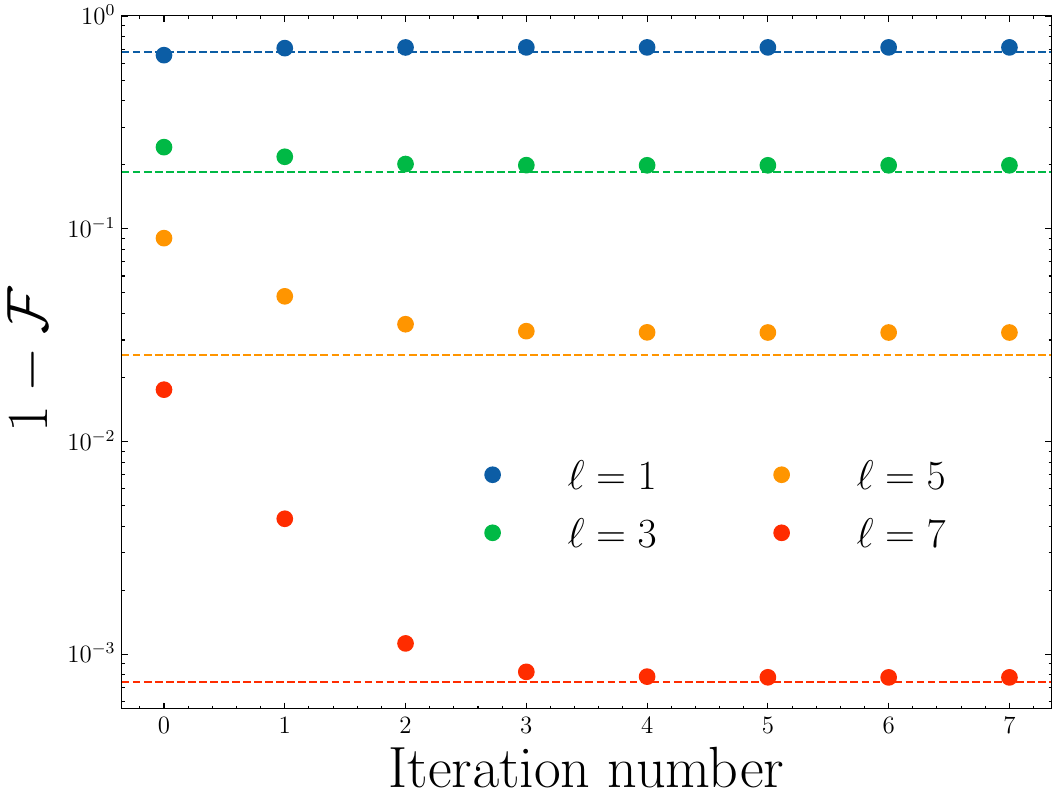}
    \caption{The fidelity of the final state after annealing using the iterative optimization procedure, \textit{without} any extra controls. At iteration zero, the evolution happens according to the infinite temperature protocol. Then, this is used to construct an ``ground state'' protocol, where the ground state is the state obtained from the previous evolution. This procedure is iterated until convergence. The dashed lines show the final state fidelity when the true ground state, obtained by exact diagonalization, is used.}
    \label{fig:zeroT-converge}
\end{figure}

Concretely, to implement this iterative procedure we first evolve the initial state according to the $T = \infty$ protocol. Then we use this approximate ground state along the protocol to recompute the action and find the new variational local AGP. Alternatively, one can perform experimental measurements of the operators entering the action $S_\lambda$ and then minimize it with respect to the variational parameters.  We note that this local  quadratic minimization can be always done on a classical computer. Then, the same initial state can be evolved according to this new protocol, which will produce a new wavefunction leading to the new action and new variational AGP. This procedure can be iterated, and provided that the orthonormal Krylov space construction of the AGP is used, the iterative scheme generally converges. We illustrate the results of this approximate ground state optimization in Figure \ref{fig:zeroT-converge}. While the converged fidelity is slightly less than that obtained using the true ground state optimization, it still shows significant improvement if large enough values of $\ell$ are used.

In this section, we have applied this iterative procedure to local CD driving without any extra controls, i.e. with a fixed ground state path. This can also be combined with the with the extra controls protocol that we describe in this work for further improvement. This motivates the development of a joint variational principle for both the variational AGP coefficients $\alpha_k$ and the extra control couplings $\beta^{(n)}$. We leave the development of this principle to future work.

%%%%%%%%%%%%%%%%%% CONCLUSION %%%%%%%%%%%%%%%%%%
\section{Conclusions \& Outlook} \label{sec:conclusions}

Counterdiabatic driving can prepare quantum states adiabatically on timescales that are much shorter than typical decoherence times. By restricting the locality of the CD driving, we can obtain experimentally-accessible protocols for approximately preparing the ground states of interesting many-body systems.

We have proposed a systematic method for finding extra control terms for quantum annealing protocols with which to augment a Hamiltonian, so that local CD driving prepares a target ground state with exponentially better fidelity. Phrased another way, our method specifies how to find the paths through the space of possible Hamiltonian couplings along which approximate CD driving will be more efficient. We find these paths by restricting the additional terms to those which preserve the commutator structure of the original CD driving term. These terms can be effectively engineered via Floquet protocols without needing to couple directly to any new terms.  We have tested this method on 1D spin chains with short- and long-range interactions and showed that it allows for a large increase in fidelity when preparing GHZ states via quantum annealing.

While this work has been principally concerned with proposing this technique and testing it with simple models, there are many others to which this recipe for more efficient paths might be applied. Going beyond one-dimensional systems, exploring models with frustration, or mappings to e.g. satisfiability problems on arbitrary graphs are all areas where this recipe may yield improvement. Another interesting direction is to pursue a joint variational principle for both the variational AGP coefficients and the extra control couplings. There are also more fundamental questions yet to be answered, such as rigorously connecting this approach and similar types of shortcuts \cite{Ibanez2012}, or similar methods in quantum dynamics such as the flow equation approach \cite{Kehrein2006}.

%%%%%%%%%%%%%%%%%%%%%%%

%%%%%%%%%%%%%%%%% ACKNOWLEDGMENTS %%%%%%%%%%%%%%%%%%

\section*{Acknowledgments}
This work was supported by NSF Grant DMR-2103658 and the AFOSR Grant FA9550-21-1-0342. The authors thank Aashish Clerk, Andrew Daley, Callum Duncan, Michael Flynn, Stefan Kehrein and Tatsuhiko Ikeda for helpful discussions. Additionally, the anonymous referees posed several interesting questions leading to further development of this work. The authors acknowledge that the computational work reported in this paper was performed on the Shared Computing Cluster administered by Boston University’s Research Computing Services. Many of the numerical computations were performed using QuSpin \cite{quspin1, quspin2}. The exact code used is available online \footnote{See https://github.com/smorawetz/CD-extra-controls.git}.

%%%%%%%%%%%%%%%%% APPENDIX %%%%%%%%%%%%%%%%%%
\appendix

\section{Construction of AGP in Krylov space} \label{appendix:krylov_construction}

Let us provide more detail about how the approximate AGP is constructed. In particular, we use a Krylov space construction of the AGP which is slightly different to that described in Ref. \onlinecite{Claeys2019}. The main  difference is that we demand that subsequent terms in the commutator expansion be orthogonal to each other. In particular, we write the AGP in terms of Krylov space operators $O_k$ with coefficients $\gamma_k$ (see also Refs.~\onlinecite{Bhattacharjee2023,Takahashi2023})

\begin{equation} \label{eqn:krylov_agp}
    A_\lambda^{(\ell)} = i \sum_{k=1}^\ell \gamma_k O_{2k-1}
\end{equation}

Before we define each of these, let us introduce the following notation for inner products and norms between operators, where we denote the operator $O$ by $O \rightarrow \vert O )$, and define the Liouvillian super-operator $\mathcal{L}$:
\begin{align*}
    (A\vert B) & = \frac{\operatorname{Tr}(A^\dagger B)}{D} \\
    \Vert A \Vert & = \sqrt{(A\vert A)}\\
    \mathcal{L} \vert O ) & = [H, O], \\
\end{align*}
where $D = 2^N$ is the dimension of the Hilbert space, and $H$ is the Hamiltonian. It is critical that in order to perform the ground state optimization discussed in \ref{subsec:finite-temp} correctly, one must replace the trace inner product by the ground state average: $(A \vert B ) = \langle \psi_0 \vert A^\dagger B \vert \psi_0 \rangle$. We reiterate that the iterative procedure described in the main text can obtain these without the use of a quantum computer. We construct the Krylov space operators $O_k$ according to the following algorithm:

\begin{algorithm}[H]
\caption{AGP Krylov space construction}\label{alg:cap}
\begin{algorithmic}
\State $A_0 \gets \partial_\lambda H$
\State $b_0 \gets \Vert A_0 \Vert$
\State $O_0 \gets A_0 / b_0$
\State $A_1 \gets \mathcal{L} \vert O_0)$
\State $b_1 \gets \Vert A_1 \Vert$
\State $O_1 \gets A_1 / b_1$
\For{$k \in 2\dots 2\cdot\ell$}
    \State $A_k \gets \mathcal{L} \vert O_{k-1}) - b_{k-1} \vert O_{k-2} )$
    \State $b_k \gets \Vert A_k \Vert$
    \State $O_k \gets A_k / b_k$
\EndFor
\end{algorithmic}
\end{algorithm}

 \noindent From this, it is apparent that the choice of Lanczos coefficients $b_k$ enforces the condition that $(O_i \vert O_j) = \delta_{ij}$.

 Now, we use the $2 \ell$ Krylov operators $O_k$ and Lanczos coefficients $b_k$ to construct the $\ell$-th order approximate AGP. To do this, we observe that the action $S_\lambda$ of Eq. \eqref{eqn:variational_action}, using the form of the AGP from Eq. \eqref{eqn:krylov_agp}:

\begin{equation*}
    S_\lambda = 1 + 2 \gamma_1 b_0 b_1 + \sum_{k=1}^\ell ( \gamma_k (b_{2k}^2 + b_{2k-1}^2) + 2 b_{2k} b_{2k+1} \gamma_k \gamma_{k+1})
\end{equation*}

\noindent which we can optimize by taking $\nabla_{\gamma_k} S_\lambda = 0$, giving the following equations

\begin{align*}
    0 & = \gamma_k A_k + \gamma_{k+1} B_k + \gamma_{k-1} B_{k-1} + b_0 b_1 \delta_{k,1} \\
    A_k & =  b_{2k-1}^2 + b_{2k}^2 \\
    B_k & = b_{2k} b_{2k+1} \\
\end{align*}

\noindent These equations can be solved recursively. The solutions are given by

\begin{align*}
    \gamma_1 & = \frac{-b_0 b_1}{A_1 - r_1 B_1} \\
    \gamma_{k+1} & = -r_k \gamma_k \\
    r_{k-1} & = \frac{B_{k-1}}{A_k - r_k B_k} \\
    r_{\ell-1} & = \frac{B_{\ell-1}}{A_\ell}
\end{align*}

In summary, at each discrete time step while solving the Schr\"odinger equation, we compute $2 \ell$ Lanczos coefficients $b_k$ and Krylov operators $O_k$, then compute $B_k$ for $k \in [1,\ell-1]$ and $A_k$ for $k \in [1,\ell]$ using the $b_k$. Then, we compute $r_k$ recursively, starting with $r_{\ell-1}$ and terminating with $r_1$. Then, we can compute $\gamma_k$ and combining this with the previously obtain Krylov operators $O_k$, we have the approximate AGP given by Eq. \eqref{eqn:krylov_agp}.

\section{Detailed Construction of AGP with Extra Controls using Floquet Pulses} \label{appendix:floquet}

\begin{figure}
    \centering
    \includegraphics[width=\linewidth]{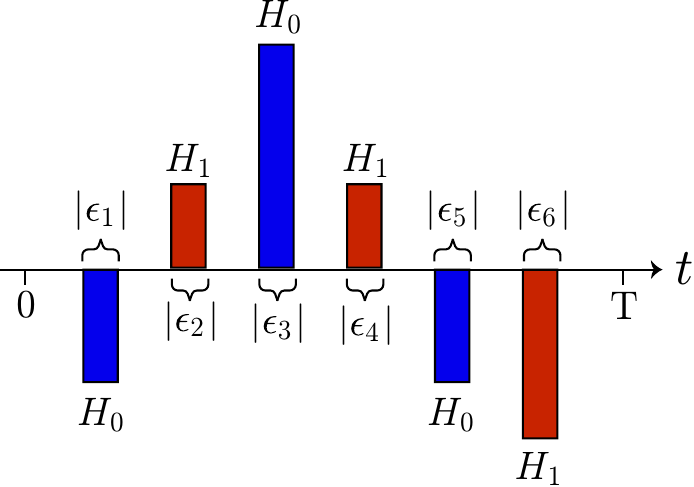}
    \caption{A cartoon illustration of the kinds of a pulse sequence required for implementation of the augmented protocol. This echo-type sequence is designed in a way that different pulses almost cancel each other such that the terms appearing in different order of the Magnus expansion are of the same order.}
    \label{fig:pulse-cartoon}
\end{figure}

In Ref.~\onlinecite{Claeys2019}, an implementation of the AGP and hence of the CD protocol via Floquet engineering is given. This is done by periodically driving both the Hamiltonian $H$ and the deformation $\partial_\lambda H$ at some high frequency $\omega$ and its odd multiples at $3\omega,\, 5\omega,\dots$. The driving amplitudes of each harmonic are then fine tuned such that the Magnus expansion of the corresponding Floquet Hamiltonian match the desired CD protocol. This Floquet protocol was, for example, recently realized in the IBM quantum simulator~\cite{Hegade_2021}. One can extend that approach to also engineer extra even commutators entering the augmented CD Hamiltonian. Rather than doing so, here we consider an alternative approach, where instead of utilizing higher order harmonics of the Floquet protocol we design an appropriate pulse sequence within the Floquet period. Then by utilizing pulse strengths as degrees of freedom we require that the emerging Floquet Hamiltonian matches the desired driving protocol. Such an approach was pioneered in the NMR literature. It is now widely used in Floquet engineering on various platforms~\cite{goldman2014, choi2020, tyler2023}. In Fig \ref{fig:pulse-cartoon} we illustrate a particular example of such a pulse sequence.

As an illustration of this method, we show here how one can engineer the pulse sequence to generate the Floquet Hamiltonian containing the desired commutators, which appear at second order in the Magnus expansion (see below). It is conceptually straightforward to generalize this procedure to higher orders or protocols with more pulses. The Magnus expansion is controlled by the Floquet period $T$, or equivalently the Trotterization time step, which should be sufficiently short. The other time scale if the protocol time $\tau\gg T$ controls the rate of change of the coupling $\dot\lambda$.

We choose the strength of the pulses so that we get the terms we want in the Magnus expansion at first and second order, and as long as the expansion is written in terms of these small parameters the higher order terms in the expansion may be neglected. Then the main goal of the protocol design is to choose the pulse strengths such that the low order terms match those in the desired CD Hamiltonian. As lower order commutator terms are parametrically larger than higher order terms in the driving period $T$. In order to make them of the same order, one needs to design echo-type pulses as is routinely done in NMR literature.

Within each period of the Floquet drive we have a sequence of alternating pulses of $H_0$ and $H_1$, where the $i$-th pulse has strength $\epsilon_i$. We want all of the pulses to be the same size, i.e. $\epsilon_i = \epsilon \eta_i$ where $\eta_i \sim O(1)$ as $T\to0$. Since we are targeting the Floquet Hamiltonian $H_F$ to second order with coefficients proportional to $T$, and the coefficients of second order terms in the Magnus expansion have three powers of the $\epsilon_i$, we take $\epsilon = T^{1/3}$.

Here we consider here a particular six-pulse sequence, with terms $\{\epsilon_1 H_0, \epsilon_2 H_1, \epsilon_3 H_0, \epsilon_4 H_1, \epsilon_5 H_0, \epsilon_6 H_1\}$. Computing the Magnus expansion~\cite{Bukov_2015} up to $O(T)$ gives:
\begin{align*}
    H_F T & = f_0 H_0 + f_1 H_1 - i f_{10} [H_1, H_0] \\
    & + f_{010} [H_0, [H_1, H_0]] + f_{110}[H_1, [H_1, H_0]]
\end{align*}
The coefficients $f_i(\lambda)$ are determined by Magnus expansion, and have been worked out in full for an arbitrary number of pulses and higher order terms \footnote{For code to calculate these, use Mathematica notebook at https://github.com/smorawetz/CD-Floquet-pulses.git}. The first few terms are
\begin{align*}
    f_0 & = \epsilon_1 + \epsilon_3 + \epsilon_5, \quad 
    f_1  =  \epsilon_2 + \epsilon_4 + \epsilon_6 \\
    f_{01} & = \epsilon_2 {\epsilon_1 + \epsilon_3 + \epsilon_5\over 2}  + \epsilon_4 {\epsilon_1 + \epsilon_3 - \epsilon_5\over 2}  + \epsilon_6 {\epsilon_1 - \epsilon_3 - \epsilon_5\over 2} \\
    & \vdots
\end{align*}

We will now match this expansion with $\tilde{H}_{\rm CD} T$. We assume that $\lambda$ changes sufficiently slowly so that it is nearly constant within a Floquet period. The CD Hamiltonian that want to match is:

\begin{align*}
    \tilde{H}_{\rm CD} = \lambda H_0 + (1-\lambda)H_1 - i \dot{\lambda} \alpha_1 [H_1, H_0] \\ + \beta^{(1)} [H_0, [H_1, H_0]] + \beta^{(2)} [H_1, [H_1, H_0]]
\end{align*}

\noindent so that the corresponding terms in the Magnus expansion are

\begin{align*}
    f_0 = \lambda, f_1 = (1-\lambda), f_{01} = \dot{\lambda} \alpha_1(\lambda) \\
    f_{010} = \beta^{(1)}(\lambda), f_{110} = \beta^{(2)}(\lambda) \\
\end{align*}

Writing the pulse strengths as $\epsilon_i = T^{1/3} \eta_i$, the following protocol can be used:
\begin{align*}
    \eta_1 & = - \frac{\beta_1^{2/3}}{\beta_2^{1/3}} + \sqrt{\frac{\alpha \beta_1}{3 \beta_2}}  \dot{\lambda}^{1/2} T^{1/6} + {1\over3}(1 - \lambda) T^{2/3} \\
    \eta_2 & = \frac{\beta_2^{2/3}}{\beta_1^{1/3}} + \sqrt{\frac{3 \alpha \beta_2}{\beta_1}} \dot{\lambda}^{1/2} T^{1/6} + {1\over3}\lambda T^{2/3} \\
    \eta_3 & = 2 \frac{\beta_1^{2/3}}{\beta_2^{1/3}} +  \sqrt{\frac{\alpha \beta_1}{3 \beta_2}} \dot{\lambda}^{1/2} T^{1/6}  + {1\over3}(1 - \lambda) T^{2/3} \\
    \eta_4 & =  \frac{\beta_2^{2/3}}{\beta_1^{1/3}} + {1\over3}\lambda T^{2/3}  \\
    \eta_5 & = - \frac{\beta_1^{2/3}}{\beta_2^{1/3}} - 2\sqrt{\frac{\alpha \beta_1}{3 \beta_2}} \dot{\lambda}^{1/2} T^{1/6} + {1\over3}(1 - \lambda) T \\
    \eta_6 & = - 2\frac{\beta_2^{2/3}}{\beta_1^{1/3}} - \sqrt{\frac{3 \alpha \beta_2}{\beta_1}} \dot{\lambda}^{1/2} T^{1/6}  + {1\over3} \lambda T^{2/3},  \\
\end{align*}

This choice of the protocol guarantees that $H_F$ and $\tilde H_{\rm CD}$ coincide up to the terms scaling as higher powers of $T$, which can be made arbitrary small by taking the limit $T\to 0$.

\section{Performance of YY vs. ZXZ controls in Short-Range Model} \label{appendix:YY_vs_ZXZ}

In the main text, we consider the preparation of GHZ states in the short-range model using two extra control Hamiltonians: $H_c^{(1)} = YY$ and $H_c^{(2)} = ZXZ$. Here we briefly analyze the optimal directions in this extra control space. We have the following annealing problem:

\begin{align*}
    \tilde{H}(\lambda) & = \lambda H_0 + (1-\lambda)H_1 + H_c(\lambda) \\
    H_0 & = -ZZ,\quad H_1 = -X \\
    H_c(\lambda) & = \beta_{YY} \sin (\pi \lambda) YY + \beta_{ZXZ} \sin(\pi \lambda) ZXZ
\end{align*}

\begin{figure}
    \centering
    \includegraphics[width=\linewidth]{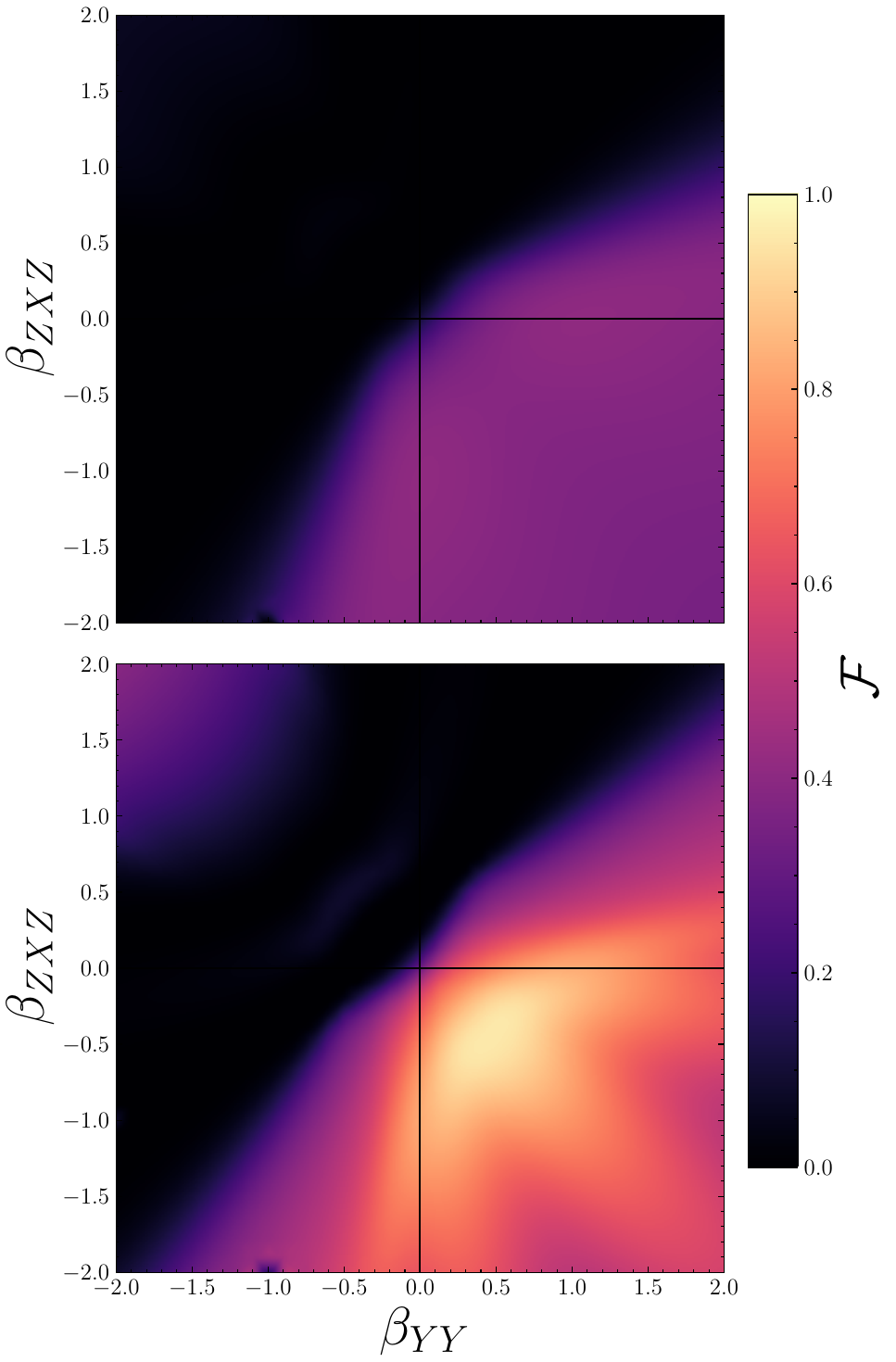}
    \caption{The fidelity obtained by augmenting the short-range GHZ state preparation for $N = 10$ by either $YY$ or $ZXZ$ controls. The origin represents the fidelity of state preparation without any extra controls. The top shows $\ell = 1$ an the bottom shows $\ell = 2$. The x and y axis show the coefficient for the $YY$ and $ZXZ$ controls respectively. From the symmetry of this plot, it is clear that neither the $YY$ or $ZXZ$ term has an advantage over the other. Starting at $\ell = 2$, using both simultaneously provides a significant advantage.}
    \label{fig:contourplot}
\end{figure}

In Figure \ref{fig:contourplot}, we show a contour plot of the final fidelity for different combinations of $\beta_{YY}$ and $\beta_{ZXZ}$ using local CD driving with $\ell = 1$ and $\ell = 2$ as defined in Eq. \eqref{eqn:commutator_ansatz} for the Hamiltonian above with $N = 10$. We find that for $\ell=1$ the optimal protocol is very close to one where only one of the controls is used, i.e. where either $\beta_{YY} = 0$ or $\beta_{ZXZ} = 0$.  The situation reverses for $\ell=2$, where the optimal performance is achieved for $\beta_{ZXZ}\approx -\beta_{YY}$. Figure \ref{fig:local_model_yy_zxz_results} in the main text shows the protocol performance along the corresponding optimal directions.

\section{Detailed Steps to Implement the Method} \label{appendix:how_to_use_summary}

In this section, we will describe in detail the step-by-step process by which we find these improved paths for local CD driving (see also Ref.~\onlinecite{Cepaite2023}). The first step is to determine the Hamiltonian for the physical system of interest, and represent it as a matrix. Furthermore, it is useful to leverage any symmetries present in the system and so construct the Hamiltonian in the relevant symmetry sector, which can be done easily with QuSpin or equivalent software package. This will make the local CD driving more efficient by reducing the number of transitions it tries to suppress. In the language of Eq. \eqref{eq:annealing}, this means determining in $H_0$ or $H_1$ and then forming
\begin{equation*}
    H(\lambda) = \lambda H_0 + (1-\lambda) H_1
\end{equation*}
\noindent This corresponds to the red path in Figure \ref{fig:path-cartoon}. The next step is to ``augment'' this Hamiltonian by adding to it two extra control Hamiltonians, taking:

\begin{equation*}
    \tilde{H}(\lambda) = H(\lambda) + \beta^{(1)} \sin(\pi \lambda) H_c^{(1)} + \beta^{(2)} \sin(\pi \lambda) H_c^{(2)}
\end{equation*}

\noindent where $H_c^{(1)}$ and $H_c^{(2)}$ are defined as in Eq. \eqref{eq:extra_even}. This corresponds to the blue path in Figure \ref{fig:path-cartoon}. We highlight that this is only the simplest possible choice such that the extra control Hamiltonians have no effect at the beginning ($\lambda = 0$) and end ($\lambda = 1$) of the protocol, and that we can even further improve the state preparation fidelity by taking further harmonics, i.e. taking $\beta^{(i)} \sin (\pi \lambda) \rightarrow \sum_k \beta_k^{(i)} \sin(k \pi \lambda)$ at the price of introducing more variational parameters. More sophisticated methods such as quantum optimal control might also be employed for even further improvement.

The next step is perform local counterdiabatic driving. Using the ``augmented'' Hamiltonian $\tilde{H}$ in the Liouvillian $\mathcal{L} = [\tilde{H}, \cdot]$, we construct the approximate AGP $\tilde{A}_\lambda^{(\ell)}$ using the procedure outlined in Section \ref{appendix:krylov_construction} of this Appendix. We add this operator to form
\begin{equation*}
    \tilde{H}_{CD}(\lambda) = \tilde{H}(\lambda) + \dot{\lambda} \tilde{A}_\lambda^{(\ell)}
\end{equation*}
With this local CD protocol $\tilde{H}_{CD}(\lambda)$, one can used standard numerical algorithms to solve the time-dependent Schr\"odinger equation to find the final state $\vert \psi_{evolved}(\vec{\beta}) \rangle$. By $\vec{\beta}$ we mean that the final state will depend on the variational $\beta$ parameters of the extra controls Hamiltonians.

\begin{figure}
    \centering
    \includegraphics[width=\linewidth]{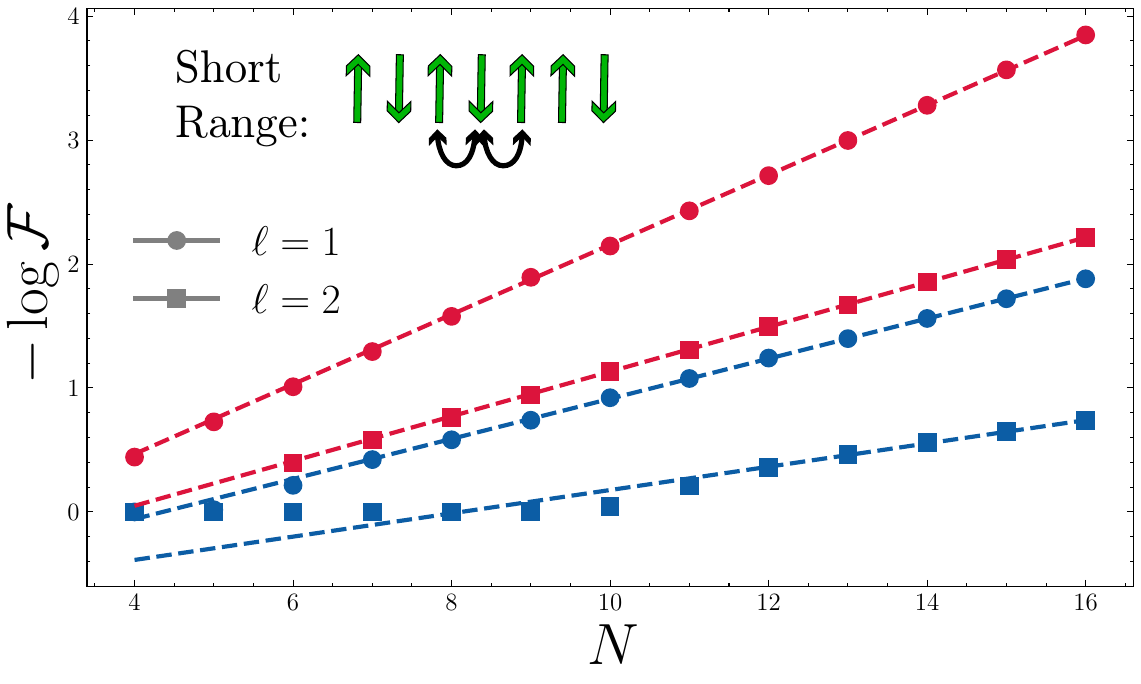}
    \caption{Improvement in the final state fidelity obtained by using the extra control ansatz $H_c^{(1)} = [H_0, [H_1, H_0]]$ and $H_c^{(2)} = [H_1, [H_1, H_0]]$ to prepare a GHZ state by annealing the short-range Hamiltonian of Eq. \eqref{eqn:short-range}. This can be compared with Figure \ref{fig:local_model_yy_zxz_results} where we use $H_c^{(1)} = YY$ and $H_c^{(2)} = ZXZ$ for the same protocol. Note that they are extremely similar because the corresponding extra controls are linear combinations of each other up to a rescaling of $ZZ$ and $X$ in the original Hamiltonian. As before, red points indicate following the ``naive'' original path, whereas blue indicates following an ``augmented'' path.}
    \label{fig:short_range_commutators}
\end{figure}

We then compute the fidelity using Eq. \eqref{eqn:fidelity}. We employ the Powell minimization algorithm \cite{powell1964} to choose the values of $\beta^{(1)}$ and $\beta^{(2)}$ which maximize the fidelity. Practically, we limit the values the coefficients can take to $\vert \beta^{(i)} \vert < 3$. This concludes the steps to implementing our protocol. In Figure \ref{fig:short_range_commutators} we apply this procedure to the short-range model Hamiltonian of Eq. \eqref{eqn:short-range}. This is the same as Figure \ref{fig:local_model_yy_zxz_results} except using the $H_c^{(1)}$ and $H_c^{(2)}$ of Eq. \ref{eq:extra_even} instead of $YY$ and $ZXZ$.

We note that this algorithm finds local, not global, minima. So while it works well for a single harmonic per extra control term, the optimization becomes more difficult for further harmonics, and a global minimizer may be required. This could be due to glassiness in the landscape of possible control protocols \cite{Day2019}.

\section{Explanation of Different Regimes of Fidelity Improvement} \label{appendix:fitting}

\begin{figure}
    \centering
    \includegraphics[width=\linewidth]{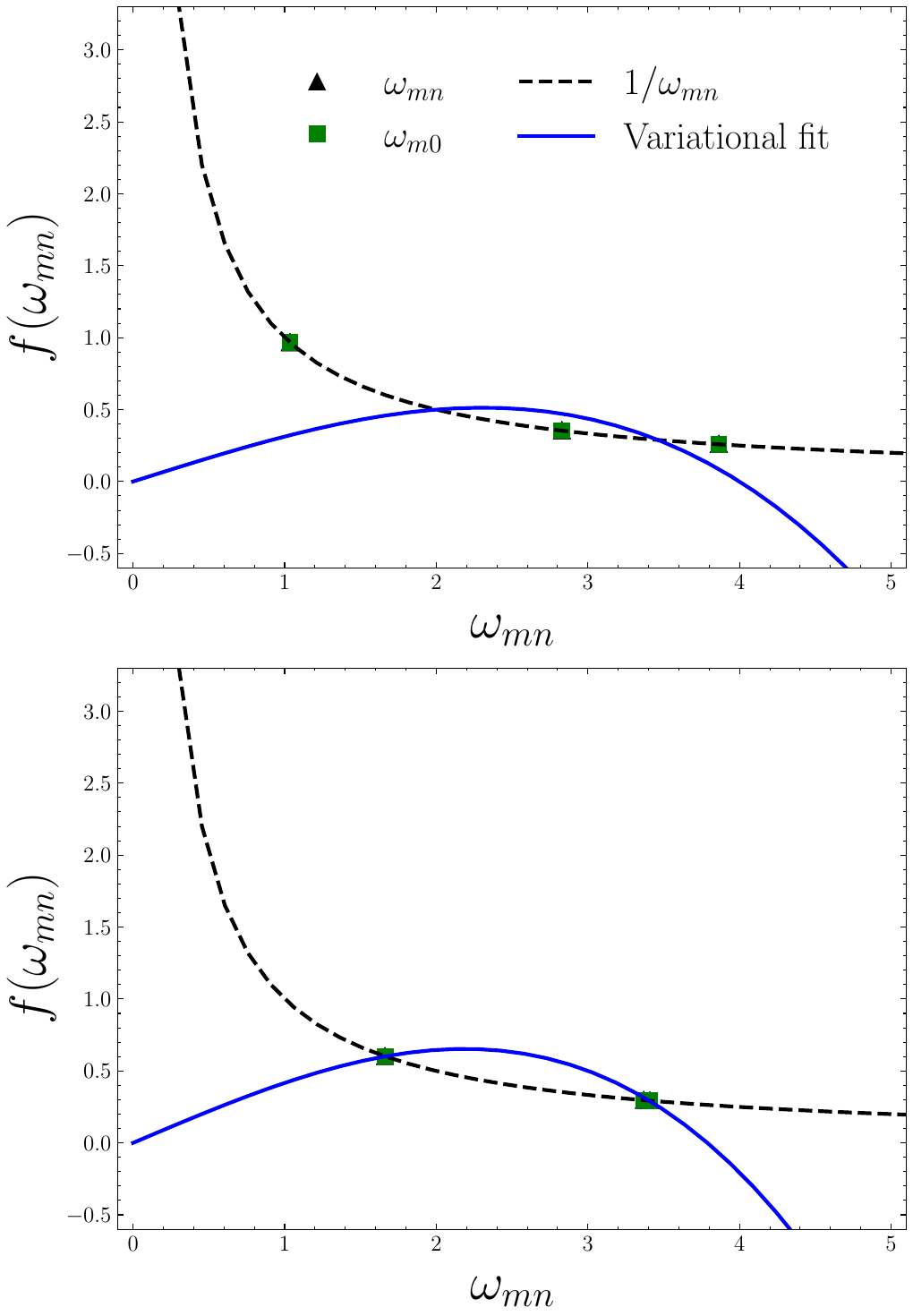}
    \caption{A snapshot of the variational optimization in terms of fitting excitation frequencies, for the short-range model with $N = 6$. On top is the ``naive path'' of the original Hamiltonian, and on the bottom is the augmented path. Although this is just one snapshot, it is apparent from this that the extra control Hamiltonian $H_c = YY$ makes local CD driving more efficient by bringing multiple excitation frequencies together.}
    \label{fig:fitting_N6}
\end{figure}

As remarked upon in the main text, in Figures \ref{fig:local_model_yy_zxz_results}, \ref{fig:LR_results} and \ref{fig:short_range_commutators}, we see two types of behavior for the fidelity of the final state prepared with extra control Hamiltonians. The first regime is a plateau where we can prepare the final state with near unit fidelity up to a certain system size, whereas the second regime has less than unit fidelity but is still exponentially improved when compared to the ``naive'' path.

To try to understand these, we refer to Ref. \onlinecite{Claeys2019}, where the variational optimization of the commutator ansatz for the AGP  (Eq. \eqref{eqn:commutator_ansatz} in this work) can be understood as simple polynomial fitting. We will summarize it here:

\begin{figure}[ht]
    \centering    \includegraphics[width=\linewidth]{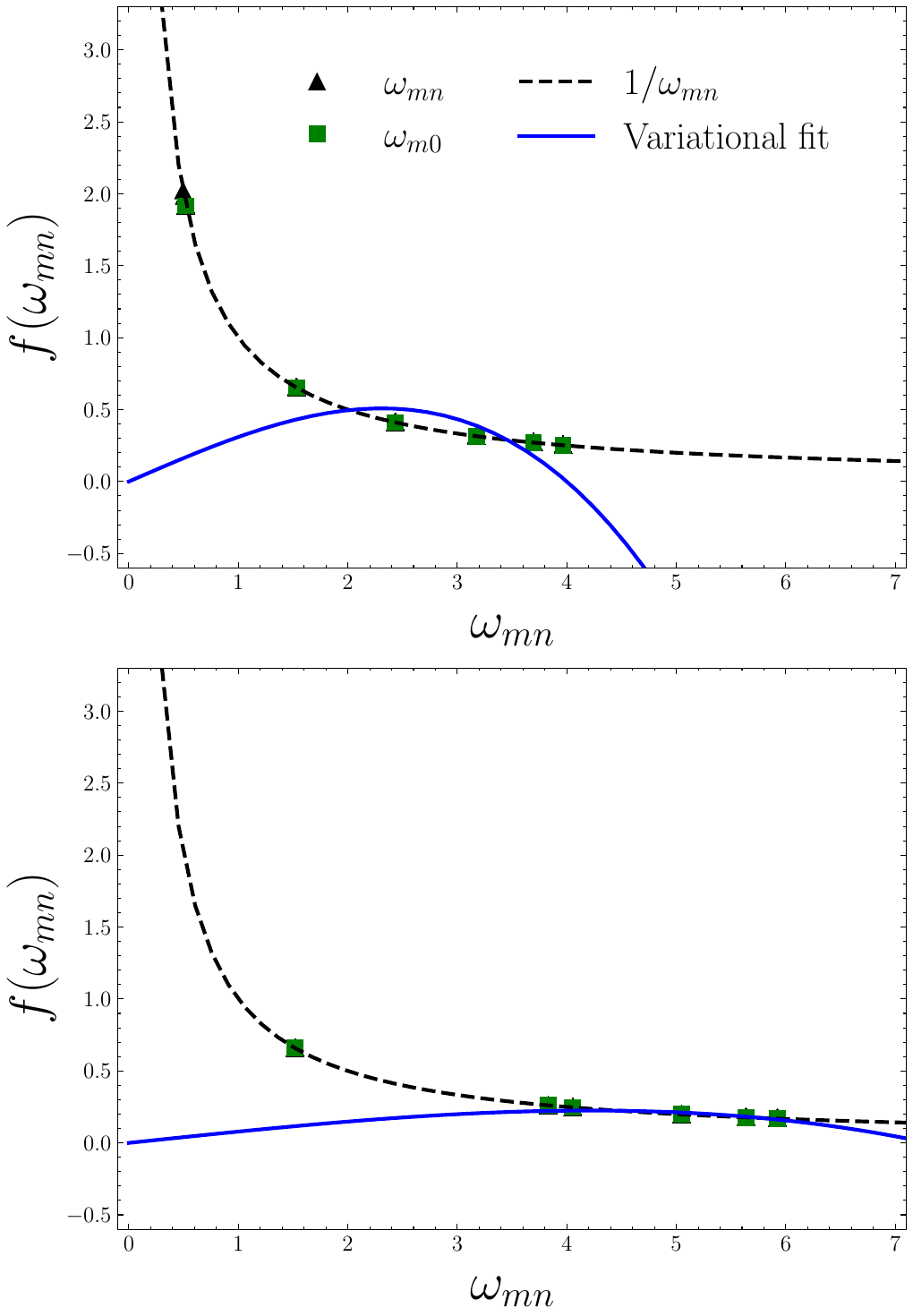}
    \caption{A snapshot of the variational optimization in terms of fitting excitation frequencies, for the short-range model with $N = 12$. The top figure is the naive path, and the bottom is the augmented path. As before, this is just one snapshot when using extra control Hamiltonian $H_c = YY$. In this instance, it makes the local CD driving more effective by shifting the excitation frequencies to larger values (increasing the gap), making them much easier to fit.}
    \label{fig:fitting_N12}
\end{figure}

The exact matrix elements of the AGP in the instantaneous energy eigenbasis satisfy
\begin{equation*}
    \braket{m | A_\lambda | n} = -i\frac{1}{\omega_{mn}} \braket{m | \partial_\lambda H | n}
\end{equation*}
where we denote $\omega_{mn} = E_m - E_n$. We can write the matrix elements of the approximate AGP $A_\lambda^{(\ell)}$ of Eq. \eqref{eqn:commutator_ansatz} as
\begin{align*}
    \braket{m | A_\lambda^{(\ell)} | n} & = i \sum_{k=1}^\ell \alpha_k \braket{m | \underbrace{[H,[H, ..., [H}_{2k-1}, \partial_\lambda H]]] | n} \\
    & = i \sum_{k=1}^\ell \alpha_k \omega_{mn}^{2k-1} \braket{m | \partial_\lambda H | n}
\end{align*}

Defining $f(\omega) = -\sum_{k=1}^\ell \alpha_k \omega^{2k-1}$, and trying to equate the matrix elements of the exact and approximate AGP gives

\begin{align*}
    f(\omega_{mn}) & = i \braket{m|A_\lambda^{(\ell)}|n}/\braket{m|\partial_\lambda H|n} \\
    \implies \omega_{mn} f(\omega_{mn})  & = \braket{m|A_\lambda^{(\ell)}|n}/\braket{m| A_\lambda|n}
\end{align*}

Therefore, if the matrix elements exactly coincide then we have that $f(\omega_{mn}) = 1/\omega_{mn}$ i.e. we have fit $\omega_{mn}$ by $f(\omega_{mn})$ for all $\omega_{mn}$ present in the system. In other words, finding the approximate AGP can be thought of as finding the best approximation to $1/\omega$ by a fixed number $(\ell)$ of odd polynomials. This fitting procedure is nicely illustrated in Figure 1 of Ref. \onlinecite{Claeys2019}. While this might seem at first glance to be a system-independent problem, it depends strongly on the range of excitation frequencies $\omega_{mn}$ which are present in the system. 

We can imagine that in a system with very few independent excitation frequencies, the approximate AGP converges very quickly due to needing to fit $1/\omega$ at only a small number of points. This is exactly what happens for small system sizes, corresponding to the unit fidelity plateaus.

This fitting procedure happens for every point in the protocol. As an illustration we analyze here the point corresponding to $\lambda = 0.5$, where the short-range model has a phase transition in the thermodynamic limit. In Figure \ref{fig:fitting_N6}, we show the polynomial $f(\omega)$ obtained for the short-range model with $N = 6$ using the variational procedure together with the exact result $1/\omega$ using $H_c = YY$. Crucially, we plot only the transition frequencies corresponding to the excitations from the ground state, which we denote by $\omega_{m0}$, which have nonzero matrix element $\langle m|\partial_\lambda H|0\rangle$. This significantly reduces the number of relevant frequencies compared to the Hilbert space size. The number of points is further reduced because there are many nearly degenerate transition frequencies. As a result the AGP is only required to suppress a small number of independent frequencies, which can be fit by a lower order polynomial perfectly.

\begin{figure*}[ht]
    \centering
    \includegraphics[width=\linewidth]{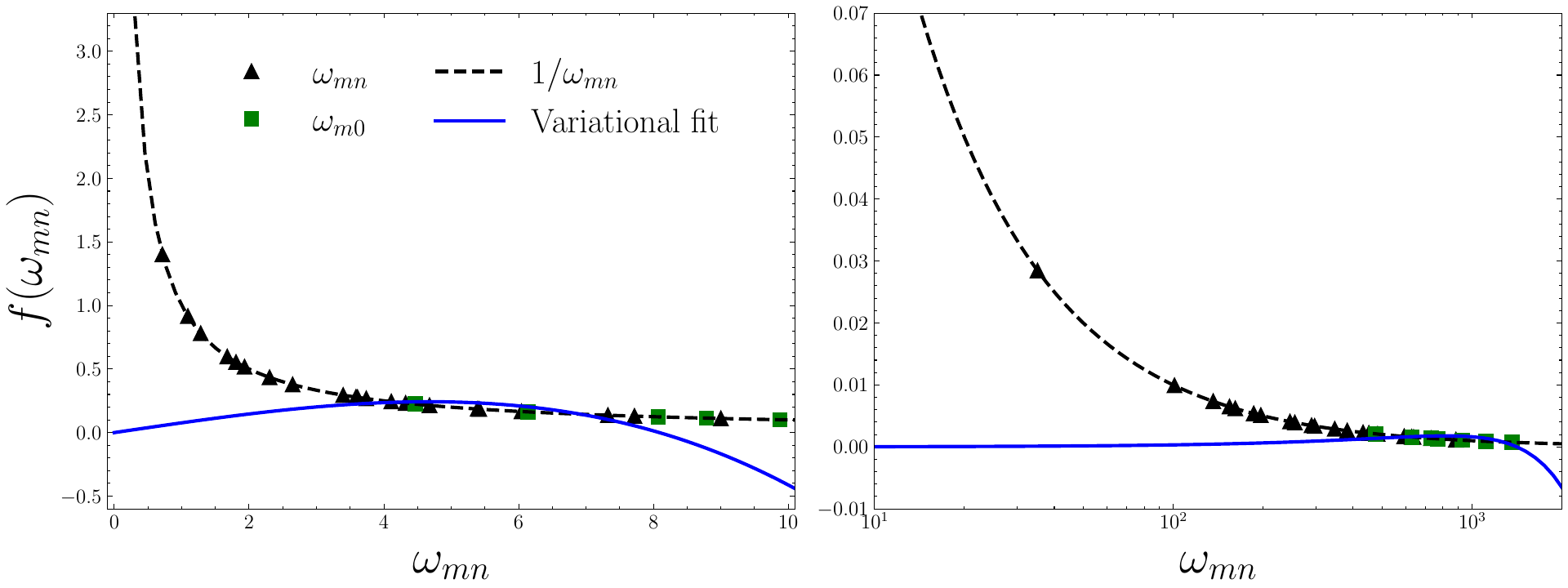}
    \caption{ The variational optimization in terms of fitting excitation frequencies, for the long-range model of Eq. \eqref{eqn:LR_ising}, with $\alpha = 2$ and $N = 6$. The long-range interactions break integrability, so the ground state is connected to far more excited states at first order in $H$. The naive path is on top, whereas the path augmented by extra controls is on the bottom. The effect of the extra controls is qualitatively similar to the integrable short-range case; they cause the frequencies to ``bunch up'' and increase the gap. Note the logarithmic scale for the augmented path.}
    \label{fig:fitting_nonintegrable}
\end{figure*}

It is apparent that the extra control Hamiltonian fulfills two tasks: i) it increases the gap in the system and ii) it leads to clustering of states such that there are fewer different frequencies $\omega_{nm}$ and thus it is easier to fit $1/\omega$ with low-order polynomials. For larger system sizes, the extra control Hamiltonian cannot group the independent excitation frequencies into just two ``clumps,'' so instead it pushes them to larger values, which is much easier to fit. This is shown in the case of $N = 12$ in Figure~\ref{fig:fitting_N12},  again with $H_c = YY$. 

The short range Hamiltonian with the $YY$ control is integrable and one might wonder if this fact allows for such efficiency of the polynomial fitting. This is, however, not the case and a very similar picture holds for nonintegrable models as well. In Figure \ref{fig:fitting_nonintegrable} we show the long-range model with $\alpha = 2$. Although there are now far more excited states which are connected to the ground state directly by $H$, qualitatively the effect of the extra control Hamiltonian is similar: it pushes frequencies to higher values and clusters them together.

%%%%%%%%%%%%%%%%%%%%%%%%%%%%%%%%%%%%%%%%%%%%%
\bibliographystyle{aipauth4-1}
\bibliography{bibliography}
%%%%%%%%%%%%%%%%%%%%%%%%%%%%%%%%%%%%%%%%%%%%%

\end{document}